\documentclass[sigconf]{acmart}

\usepackage[english]{babel}
\usepackage{blindtext}
\usepackage[nolist]{acronym}
\renewcommand\footnotetextcopyrightpermission[1]{} 
\setcopyright{none}

\acmDOI{}

\acmISBN{}

\acmPrice{}

\begin{document}
\title{Modular and Integrated AI Control Framework across Fiber and Wireless Networks for 6G}


\author{Merim Dzaferagic, Marco Ruffini, Daniel Kilper}
\affiliation{%
  \institution{Trinity College Dublin, Ireland}
}

\renewcommand{\shortauthors}{Dzaferagic M. et al.}

\begin{abstract}
The rapid evolution of communication networks towards 6G increasingly incorporates advanced AI-driven controls across various network segments to achieve intelligent, zero-touch operation. This paper proposes a comprehensive and modular framework for AI controllers, designed to be highly flexible and adaptable for use across both fiber optical and radio networks. Building on the principles established by the O-RAN Alliance for near-Real-Time RAN Intelligent Controllers (near-RT RICs), our framework extends this AI-driven control into the optical domain. Our approach addresses the critical need for a unified AI control framework across diverse network transport technologies and domains, enabling the development of intelligent, automated, and scalable 6G networks. 
\end{abstract}

\maketitle

\section{Introduction}
In this paper, we propose a comprehensive framework designed to serve as the blueprint for AI controllers in future 6G networks. Our objective is to create a modular and flexible system that can be seamlessly deployed across various network segments and devices. This framework aims to facilitate the integration of AI-driven controls into the entire network infrastructure, enabling intelligent, automated management and optimization.

The motivation behind this work stems from the critical role AI will play in the evolution of communication networks. While the O-RAN community has made significant strides in developing a control framework for the Radio Access Network (RAN), our goal is to extend this concept to encompass the entire network, including transport, core, and optical domains. We envision a \ac{dmmai} framework that defines the essential building blocks for its operation, yet remains flexible enough to accommodate diverse designs and ensure interoperability across different network elements.

Our approach involves establishing a modular design consisting of five key components: an AI engine, a registry, two message brokers, and a protocol translation module. Additionally, we introduce a Node control module that can be deployed on all network nodes, enabling them to communicate with the controller and accept control directives. This design ensures that the framework can be adapted to various network configurations and requirements.

The main contribution of our work includes providing a detailed blueprint for an AI controller that supports decentralized AI control across all parts of a communication network. Our modular framework facilitates the integration of AI capabilities, promoting intelligent, zero-touch networks that can self-configure, monitor, and repair at scale. This research lays the groundwork for future development, validation, and standardization of AI-driven control mechanisms in large-scale 6G networks.

\section{Related Work}
The landscape of \ac{ai}-driven control in communication networks has seen significant advancements, particularly within the context of the Open Radio Access Network (O-RAN). The O-RAN Alliance has introduced the concept of the near-real-time RAN Intelligent Controller (near-RT RIC), which plays a pivotal role in enabling intelligent, software-defined management of \acp{ran} by supporting multi-party applications for enhanced performance and flexibility~\cite{polese2023understanding}. Lightweight control applications (i.e. xApps and rApps) have been developed for open-source \acp{ric} \cite{lee2020hosting, xavier2023machine}. Integration of \ac{ric}-based AI network controls with \ac{sdn} is also being explored \cite{xavier2024cross}. While the O-RAN architecture provides a standardized platform for \ac{ai}/\ac{ml} control in radio networks, there is no equivalent for transport and optical networks~\cite{lee2020hosting}. These domains use \ac{sdn} platforms like \ac{onos} to orchestrate network functions \cite{akinrintoyo2023reconfigurable}. In optical networks, \ac{ai} control focuses on offline \ac{qot} estimation, wavelength routing, and fault management \cite{mata2018artificial}. Extensions of \ac{onos} for online \ac{ai} and hierarchical control have been explored \cite{morro2018automated}. Applying \ac{ai} to radio and optical network control is relatively new, lacking accepted cross-domain frameworks, which will be essential for end-to-end AI controls. 



The evolution of the Cloud-RAN has also contributed to network performance enhancements. Defined by the \ac{ngnm} alliance, Cloud-RAN's architecture includes the \ac{cu}, \ac{du}, and \ac{ru}, connected via the Low Layer Split transport interface or fronthaul~\cite{das2019variable}. Traditional dedicated fiber fronthaul networks are being reconsidered in favor of \ac{pon} to reduce connectivity costs and support the increased small cell densification required for 5G and beyond. However, the challenge of latency caused by \ac{dba} remains a significant issue, particularly due to the lack of coordination between PON and DU upstream schedulers~\cite{slyne2024demonstration}.

The concept of "Cooperative DBA" was introduced to mitigate this latency by facilitating better coordination between optical and mobile networks. This concept has been further developed and standardized as the "Cooperative Transport Interface (CTI)," which allows for enhanced synchronization between PON and RAN upstream schedulers, resulting in more efficient resource allocation and reduced latency~\cite{nomura2017first}. Despite these advancements, the integration of AI capabilities into the \ac{pon} has not been fully explored.

Our work aims to bridge this gap by extending the principles of O-RAN near-RT RIC controllers into the optical domain, thereby promoting a unified AI-driven control framework across both wireless and optical networks. By adding AI capabilities to the \ac{pon}, we aim to achieve seamless integration and coordination between these domains, enabling more sophisticated and intelligent network management. This approach not only leverages the existing strengths of the O-RAN framework but also enhances it with AI-driven insights, facilitating the creation of intelligent, zero-touch networks that can self-configure, monitor, and repair across a broad range of network scenarios.


\section{System Design}
\begin{figure}[tp]
\centering
\includegraphics[scale=0.22]{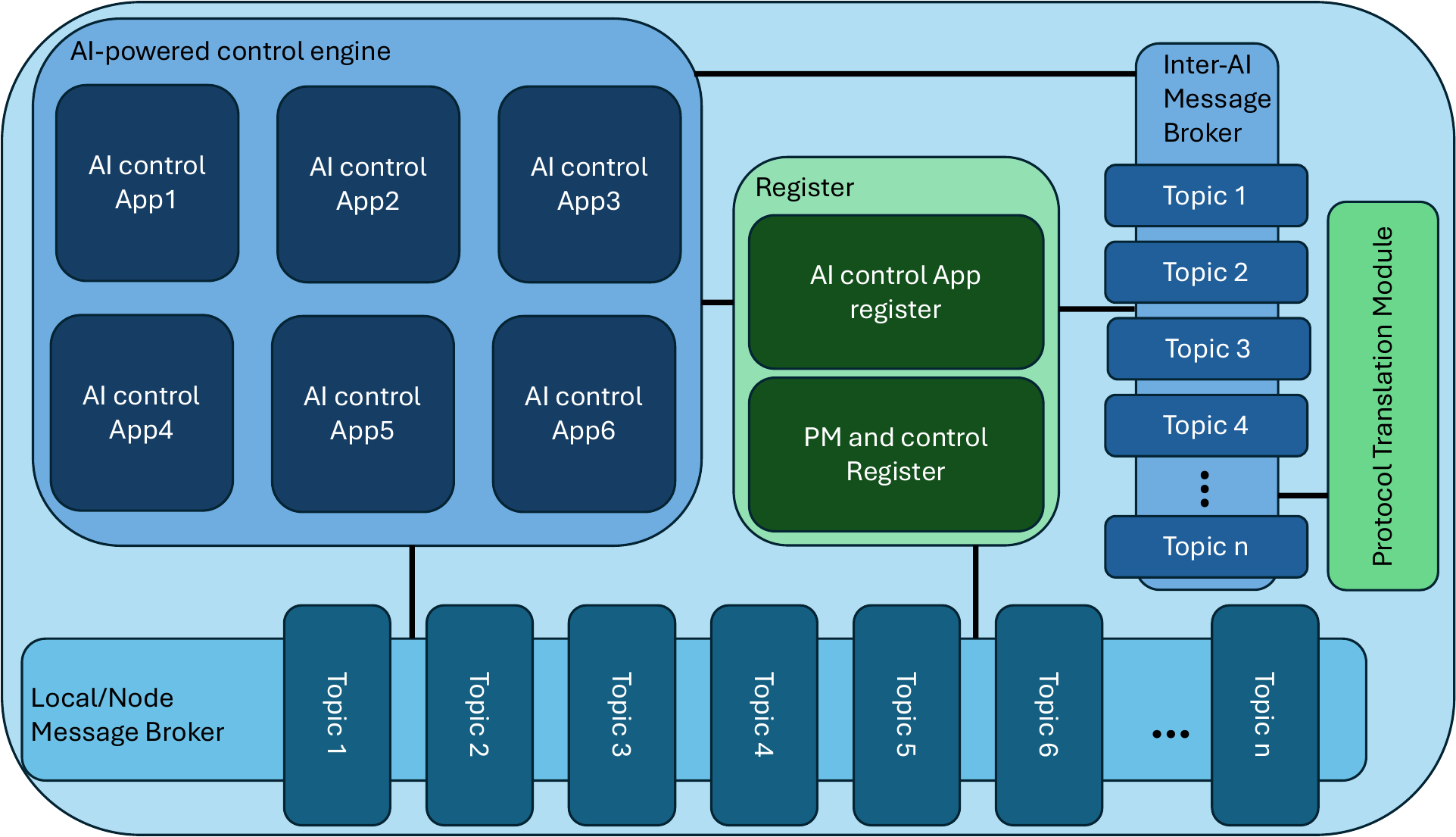}
\caption{System Architecture}\label{fig:system_architecture}\vspace{-2em}
\end{figure}

The framework implementation, illustrated in Figure~\ref{fig:system_architecture}, consists of five modules: two Message Brokers, an AI-Powered Control Engine, a Register, and a Protocol Translation Module. This modular design provides the flexibility needed to adjust the implementation to the specific requirements of a network node or network domain.

\textbf{Message Brokers:}
There are two types of message brokers in our design: the Local/Node Message Broker and the Inter-AI Message Broker. Both brokers are based on a publish-subscribe design pattern, which enhances flexibility by decoupling publishers and subscribers. This decoupling allows us to independently scale the number of AI control applications and the managed nodes.

\textit{The Local/Node Message Broker} facilitates communication between the controller and the managed nodes, handling north-south communication. Each topic within this broker is dedicated to either performance measurements published by network nodes or control messages sent from AI control applications to the network nodes. \textit{The Inter-AI Message Broker} supports east-west communication, enabling interaction between different AI controllers.

\textbf{AI-Powered Control Engine:}
The AI-powered control engine provides an environment for running AI control applications. It manages communication through the two message brokers and ensures the correct operation of AI applications. This engine registers AI applications and retrieves information about available performance measurements and control parameters from the Register.

\textbf{Register:}
The Register stores detailed information about managed nodes, including available control parameters and exposed performance metrics. It also maintains information about AI applications, aiding in conflict resolution and enabling east-west communication (i.e. inter-AI controller communication).

\textbf{Protocol Translation Module:}
This module facilitates east-west communication between our proposed architecture and other AI controllers with different designs by performing protocol translation and adaptation. This modular design supports the integration of various controllers within the network. Direct communication between instances of our controllers bypasses the protocol translation module, occurring directly between the AI-powered control engine of one controller and the Inter-AI Message Broker of another controller.

\textbf{Node Control Module:}
A small Node Control Module must be deployed on the managed nodes. This module is a straightforward publish-subscribe module that exposes all node performance measurements to the AI controller and subscribes to topics on the Local/Node Message Broker to receive control messages intended for the specific node.
\vspace{-1em}

\subsection{Workflow}
\begin{figure}[tp]
\centering
\includegraphics[trim=0 400 0 0,clip,scale=0.35]{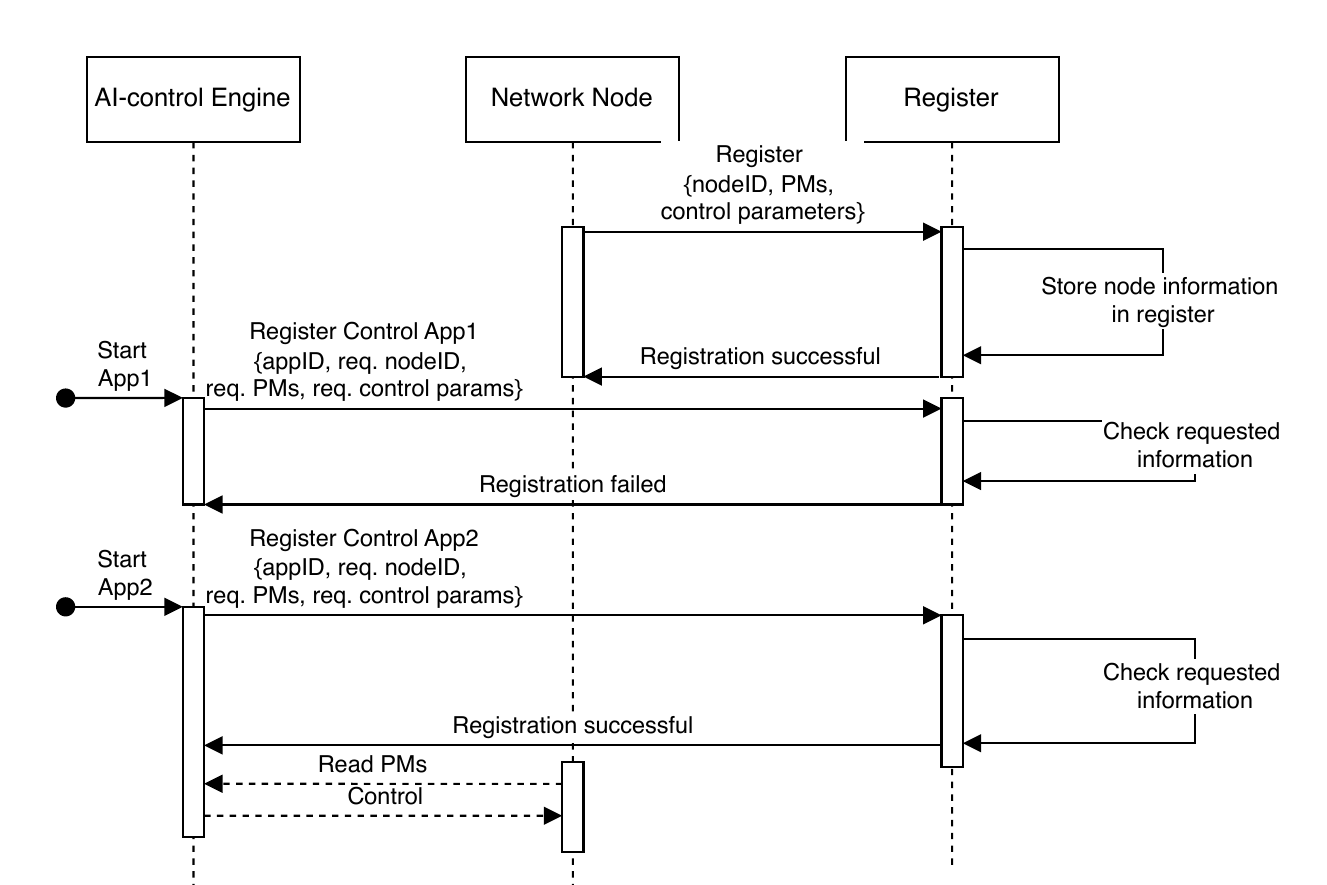}
\caption{Node and AI control application registration and control workflow.}\label{fig:registration_and_control}\vspace{-1.2em}
\end{figure}

Figure~\ref{fig:registration_and_control} depicts the workflow for running the AI controller with managed nodes. The first step in the process is the Node Registration. When a node is activated, it registers with the AI controller. This involves an initial message exchange with the Register to store the node ID along with a list of available performance measurements and control parameters.

The second step involves the AI Control Application Registration. Once nodes are registered, AI control applications can begin their registration process. During registration, the AI control engine exchanges information with the Register to determine the required performance measurements and intended use of control parameters. If the necessary measurements or control parameters are unavailable for a specific node, the Register sends an exception to the AI control engine, causing the registration to fail. If all required elements are available, the registration succeeds, and information about the control parameters and the AI control application is stored in the Register.

Once the registration of the network nodes and the AI control applications is successful, the AI control applications will have access to the exposed performance measurements and will have the capability to control the controllable parameters of the network nodes. This access allows the AI applications to dynamically monitor and adjust network performance, ensuring responsiveness to changing network conditions.

Decentralized Collaboration: If multiple controllers need to collaborate in a decentralized manner, this is enabled through the Inter-AI Message Brokers, allowing seamless interaction and coordination between different AI controllers.

Our previous work, published in \cite{slyne2024demonstration}, demonstrates various elements of the proposed framework facilitating communication between \ac{ran} elements and the \ac{pon}, enabling dynamic resource scheduling in the \ac{pon} to minimize latency in the \ac{ran} fronthaul. Although this work does not incorporate AI-powered control, the framework's flexibility allows for the introduction of AI capabilities to the \ac{pon}. The ability to decouple and use different elements of the framework separately highlights its adaptability, enabling cooperation and \ac{ai} functionalities in different network domains. 

\section{Conclusion}
In this paper, we have proposed a comprehensive framework designed to serve as a blueprint for \ac{ai} controllers in future 6G networks. The modular design allows for flexible and scalable implementation tailored to the specific needs of various network nodes and domains. By decoupling publishers and subscribers through a publish-subscribe pattern, our system ensures efficient communication and control across different segments of the network.

{\color{black} One of the key contributions of our framework is its ability to extend the principles of O-RAN near-RT RIC controllers into the optical domain. Integrating AI capabilities into the \ac{pon} enhances functionality, enabling seamless coordination and resource management between optical and wireless networks. This supports decentralized AI-driven control for intelligent, automated network management. Adding \ac{ai} capabilities to the \ac{pon} enables use-cases that involve predictive capacity management, anticipating \ac{ran} usage, and, in a multi-channel (\ac{twdm} \ac{pon}) scenario, reassigning \ac{onu}-\acp{ran} and other \acp{onu} to different channels to optimize \ac{qos}. This would enhance wavelength management and optimally distributes \ac{ran} traffic across channels, meeting high-priority, low-latency needs.}


Overall, the proposed framework represents a significant step toward the realization of intelligent, zero-touch networks capable of self-configuration, monitoring, and repair at scale. By bridging the gap between wireless and optical domains and enabling decentralized AI control, our design enables advanced, AI-driven communication networks that meet the demanding requirements of future 6G applications.


\begin{acronym}
  \acro{ai}[AI]{Artificial Intelligence}
  \acro{aic}[AIC]{\ac{ai} Controller}
  \acro{dmmai}[DMMAI]{Decentralized Multi-party, Multi-network AI}
  \acro{pon}[PON]{Passive Optical Networks}
  \acro{dba}[DBA]{Dynamic Bandwidth Allocation}
  \acro{twdm}[TWDM]{Time and Wavelength Division Multiplexing}
  \acro{onu}[ONU]{Optical Network Unit}
  \acro{qos}[QoS]{Quality of Service}
  \acro{ngnm}[NGNM]{Next Generation Mobile Networks}
  \acro{sdn}[SDN]{Software-Defined Networking}
  \acro{onos}[ONOS]{Open Network Operating System}
  \acro{nos}[NOS]{Network Operating System}
  \acro{qot}[QoT]{Quality of Transmission}
  \acro{tip}[TIP]{Telecom Infra Project}
  \acro{ml}[ML]{Machine Learning}
  \acro{ran}[RAN]{Radio Access Network}
  \acro{rt}[RT]{Real Time}
  \acro{ric}[RIC]{\acs{ran} Intelligent Controller}
  \acro{bs}[BS]{Base Station}
  \acro{ue}[UE]{User Equipment}
  \acro{du}[DU]{Distributed Unit}
  \acro{ru}[RU]{Radio Unit}
  \acro{dns}[DNS]{Domain Name System}
  \acro{cu}[CU]{Centralized Unit}
  \acro{zsm}[ZSM]{Zero touch network and service management}
  \acro{pai}[PAI]{Pervasive \ac{ai}}
  \acro{uav}[UAV]{Unmanned Aerial Vehicle}
  \acro{meao}[MEAO]{Multi-Access Edge Computing Application Orchestrator}
  \acro{mec}[MEC]{Multi-Access Edge Computing}
  \acro{mno}[MNO]{Mobile Network Operator}
  \acro{aiaas}[AIaaS]{\ac{ai} as a Service}
  \acro{ns}[N-S]{North-South}
  \acro{ew}[E-W]{East-West}
\end{acronym}
\bibliographystyle{ACM-Reference-Format}
\bibliography{reference}

\end{document}